\documentclass[12pt]{article}

\usepackage[a4paper,left=2.54cm,right=2.54cm,top=2.54cm,bottom=2.54cm]{geometry}
\usepackage[T1]{fontenc}
\usepackage[utf8]{inputenc}
\usepackage{lmodern}
\usepackage{parskip}
\usepackage{hyperref}
\usepackage{graphicx}
\usepackage{adjustbox}
\usepackage[square,numbers,sort&compress]{natbib}
\usepackage[dvipsnames]{xcolor}
\usepackage{authblk} 
\usepackage{float}   
\usepackage{booktabs} 
\usepackage{caption}

\title{\textbf{Assessing the Pedagogical Readiness of Large Language Models as AI Tutors in Low-Resource Contexts: A Case Study of Nepal's K-10 Curriculum}}

\author{
Pratyush Acharya,
Prasansha Bharati,
Yokibha Chapagain,
Isha Sharma Gauli,
Kiran Parajuli
}

\date{\today}

\begin{document}

\maketitle

\section{Abstract}
The integration of Large Language Models (LLMs) into educational ecosystems promises to democratize access to personalized tutoring, yet the readiness of these systems for deployment in non-Western, low-resource contexts remains critically under-examined. This study presents a systematic evaluation of four state-of-the-art LLMs—GPT-4o, Claude Sonnet 4, Qwen3-235B, and Kimi K2—assessing their capacity to function as AI tutors within the specific curricular and cultural framework of Nepal’s Grade 5–10 Science and Mathematics education. We introduce a novel, curriculum-aligned benchmark and a fine-grained evaluation framework inspired by the "natural language unit tests" paradigm, decomposing pedagogical efficacy into seven binary metrics: Prompt Alignment, Factual Correctness, Clarity, Contextual Relevance, Engagement, Harmful Content Avoidance, and Solution Accuracy. Our results reveal a stark "curriculum-alignment gap." While frontier models (GPT-4o, Claude Sonnet 4) achieve high aggregate reliability (\textasciitilde 97\%), significant deficiencies persist in pedagogical clarity and cultural contextualization. We identify two pervasive failure modes: the "Expert's Curse," where models solve complex problems but fail to explain them clearly to novices, and the "Foundational Fallacy," where performance paradoxically degrades on simpler, lower-grade material due to an inability to adapt to younger learners' cognitive constraints. Furthermore, regional models like Kimi K2 exhibit a "Contextual Blindspot," failing to provide culturally relevant examples in over 20\% of interactions. These findings suggest that off-the-shelf LLMs are not yet ready for autonomous deployment in Nepalese classrooms. We propose a "human-in-the-loop" deployment strategy and offer a methodological blueprint for curriculum-specific fine-tuning to align global AI capabilities with local educational needs.

\section{Introduction}

\subsection{The Paradigm Shift in Educational Artificial Intelligence}
The rapid advancement of generative artificial intelligence, specifically Large Language Models (LLMs), has precipitated a paradigm shift in the global education sector. These models, trained on vast corpora of human knowledge, offer the potential to function as scalable, Intelligent Tutoring Systems (ITS) capable of delivering personalized instruction, adaptive feedback, and on-demand content generation \cite{smith2026ai, vanlehn2011its, unesco2021ai}. Unlike traditional rule-based ITS, LLMs possess the generative capacity to engage in open-ended dialogue, potentially simulating the Socratic method and fostering deeper conceptual understanding \cite{lmunit}. For developing nations, where teacher shortages and resource constraints are endemic, the promise of an ``AI tutor for every child'' represents a potential mechanism to leapfrog systemic barriers and accelerate progress toward Sustainable Development Goal 4 (Quality Education). This aspiration is rooted in Bloom's seminal ``2 Sigma Problem,'' which demonstrated that one-on-one tutoring produces learning gains two standard deviations above conventional classroom instruction \cite{bloom1984two}.

However, the deployment of such powerful technologies is fraught with pedagogical and ethical perils. The very plasticity that allows LLMs to generate creative explanations also renders them prone to "hallucinations," or the fabrication of plausible but incorrect information \cite{gpt4o_card}. Furthermore, the ease with which these models can generate complete solutions raises profound concerns regarding academic integrity and the potential atrophy of critical thinking skills among students \cite{superficial_learning}. As educational institutions worldwide grapple with these dualities, the need for rigorous, context-aware evaluation frameworks becomes paramount. We must move beyond asking whether LLMs can pass standardized tests to asking whether they can effectively \textit{teach} the material contained within those tests to a diverse student body.

\subsection{The Curriculum and Context Alignment Gap}
A critical, yet often overlooked, limitation of current LLM research is the "curriculum and context alignment gap." State-of-the-art models are predominantly trained on English-centric corpora reflecting Western cultural norms, pedagogical styles, and examples \cite{fluent_foreign}. While these models demonstrate impressive general knowledge, their ability to navigate the specific curricular requirements and cultural contexts of non-Western education systems remains largely unverified.

In the context of Nepal, this alignment gap is not merely a theoretical concern but a practical barrier to effectiveness. An AI tutor explaining the concept of "monsoon" using examples of Atlantic hurricanes, or illustrating arithmetic using dollars instead of Nepalese Rupees, introduces unnecessary cognitive load and alienation. Effective pedagogy relies heavily on ``cultural scaffolding''—connecting new concepts to the learner's lived experience, a principle grounded in Vygotsky's Zone of Proximal Development theory \cite{vygotsky1978zpd}. If an LLM fails to utilize local landmarks, indigenous flora, or culturally relevant analogies, its pedagogical utility is significantly diminished, regardless of its raw factual accuracy. This study posits that the "alignment gap" is a measurable metric of failure in current off-the-shelf models when applied to the Global South.

\subsection{The Socio-Technical Context of Nepal}
Nepal presents a unique case study for the evaluation of AI in education. The Government of Nepal has articulated a clear strategic vision for digital transformation through the \textit{Digital Nepal Framework} and the recently approved \textit{National AI Policy 2082} (2025) \cite{policy_framework}. These policy documents outline an ambition to leverage AI for socio-economic development, with a specific focus on modernizing the education sector.

However, the on-the-ground reality is characterized by a "paradox of readiness" \cite{unesco_southasia2024}. Recent empirical studies indicate high rates of AI tool experimentation among faculty but low perceived utility for core teaching tasks, with teachers showing varying degrees of familiarity and acceptance \cite{gardan2025adopting}. This discrepancy suggests that while the appetite for technology exists, the available tools—primarily generic, Western-optimized LLMs—are not meeting the specific needs of the local ecosystem. Systemic challenges, including outdated syllabi, limited digital infrastructure in rural areas, and a lack of localized digital content, further complicate integration \cite{unesco_southasia2024, digital_divide2024}. Evaluating AI readiness in this context requires a framework that acknowledges these constraints and prioritizes models that can function effectively as supportive tools for human teachers, rather than autonomous replacements.

\subsection{Research Objectives}
This research aims to bridge the gap between high-level policy ambitions and the practical realities of classroom instruction in Nepal. We are guided by the following primary research questions:
\begin{enumerate}
  \item \textbf{Pedagogical Efficacy:} How accurately and reliably do leading LLMs (GPT-4o, Claude Sonnet 4, Qwen3, Kimi K2) perform on the specific learning objectives of Nepal’s Grade 5–10 Science and Mathematics curriculum?
  \item \textbf{Contextual Intelligence:} To what extent are these models capable of generating explanations and examples that are culturally, geographically, and socially relevant to Nepalese students?
  \item \textbf{Failure Analysis:} What are the specific failure modes of these models? Do they struggle more with content complexity or pedagogical clarity?
  \item \textbf{Readiness Assessment:} Based on a fine-grained evaluation, are general-purpose LLMs ready for deployment as AI tutors in Nepal, and what interventions are necessary to close the readiness gap?
\end{enumerate}

\section{Related Work}

\subsection{Large Language Models in Education}
The literature on LLMs in education has expanded rapidly since the release of ChatGPT. Early studies focused on the models' performance on standardized tests, with results indicating capabilities exceeding human baselines in subjects ranging from law to medicine \cite{gpt4o_card}. In the domain of STEM education, research has demonstrated the potential of LLMs to function as "socratic tutors," guiding students through multi-step physics and math problems rather than simply providing answers \cite{smith2026ai}.

However, the effectiveness of these interventions is highly sensitive to the quality of the interaction. Studies utilizing the "protégé effect"—where students teach the AI—have shown promise in deepening student understanding, but rely on the AI's ability to simulate a novice learner accurately \cite{superficial_learning}. Conversely, the use of LLMs for automated feedback generation has yielded mixed results; while models can identify syntax errors in programming assignments with high accuracy \cite{frontiers_aihighed2024}, their ability to provide conceptual feedback that aids learning without revealing the solution is less robust \cite{frontiers_aihighed2024}. This literature underscores the necessity of evaluating LLMs not just as knowledge bases, but as pedagogical agents with distinct instructional behaviors.

\subsection{Evaluation Methodologies: From General Benchmarks to Unit Tests}
As the application of LLMs becomes more specialized, traditional evaluation metrics like BLEU and ROUGE have proven insufficient. These n-gram-based metrics fail to capture the semantic nuance, factual accuracy, and pedagogical utility of a response. Consequently, the field has moved toward model-based evaluation and fine-grained rubrics.

The "natural language unit tests" paradigm, formalized by Saad-Falcon et al. (2025) in their \textit{LMUnit} framework, represents a significant methodological advance \cite{lmunit}. By decomposing a complex construct like "response quality" into a series of binary, testable criteria (e.g., "Does the response contain a code snippet?" "Is the tone formal?"), this approach increases the reliability and interpretability of evaluations. Our study adapts this methodology, translating software engineering principles into pedagogical metrics (e.g., "Is the language grade-appropriate?").

Similarly, the \textit{E-EVAL} benchmark \cite{aclanthology2025} for Chinese K-12 education demonstrates the importance of localized evaluation datasets. The authors found that models optimized for English often underperform on Chinese curriculum-specific tasks, validating the need for national-level benchmarks. Our work parallels this effort, establishing the first such benchmark for the Nepalese context.

\subsection{Cultural Alignment and Bias in AI}
The challenge of cultural alignment in AI is a growing area of inquiry. Research has consistently shown that "universal" models often exhibit a strong Western bias in their values, examples, and social reasoning \cite{values_rag}. This "values gap" can lead to alienation in non-Western users. For example, Agarwal et al. (2025) demonstrated that even "regional" models trained on local languages often fail to reflect local cultural values, instead mimicking the Western norms present in their base foundation models \cite{fluent_foreign}.

In the context of education, this bias manifests as a failure of relevance. An AI tutor that uses examples from baseball or American history to explain universal concepts may fail to engage a student in South Asia. This failure is not merely cosmetic; it hinders the cognitive process of linking new information to prior knowledge. Our study explicitly quantifies this "Contextual Blindspot," contributing to the broader literature on AI fairness and inclusivity in the Global South \cite{ai_ethics_global_south}.

\subsection{The "Foundational Fallacy" and "Expert's Curse"}
Educational theory provides two critical concepts for interpreting LLM failures. The "Foundational Fallacy," described by Osborne (2007), refers to the misconception that teaching science requires a rigid, brick-by-brick assembly of facts, often ignoring the need to engage students' curiosity or simplify complex ideas without losing accuracy \cite{osborne2007science}. This connects to Piaget's constructivist theory, which emphasizes that learners actively construct knowledge through interaction with their environment \cite{piaget1970constructivism}. In the context of AI, we reinterpret this as the assumption that an AI capable of solving graduate-level physics will naturally be competent at teaching 5th-grade science.

Closely related is the "Expert's Curse" (or Curse of Knowledge), a cognitive bias where an expert assumes that the learner shares their background knowledge, leading to explanations that are technically correct but pedagogically opaque. This phenomenon is formalized in the ``Expertise Reversal Effect,'' which demonstrates that instructional techniques effective for novices can become redundant or even counterproductive for advanced learners \cite{kalyuga2003expertise}. Previous research has identified this as a potential risk in AI tutoring, where models may default to high-level academic language that alienates novice learners \cite{zerkouk2025its, nature2025its}. Our evaluation specifically investigates whether LLMs exhibit these human-like pedagogical failures.

\section{Methodology}

\subsection{The Pedagogical Evaluation Framework}
To provide a granular assessment of LLM performance, we developed a 7-metric evaluation framework based on the "natural language unit tests" paradigm \cite{lmunit}. Unlike holistic scoring, which can obscure specific weaknesses, this framework utilizes binary (Pass/Fail) criteria for each metric, ensuring high inter-rater reliability and actionable diagnostics. The metrics are defined in Table 1.

\begin{table}[H]
\centering
\caption{Pedagogical Evaluation Framework Metrics}
\label{tab:metrics}
\begin{adjustbox}{width=\textwidth}
\begin{tabular}{@{}llp{8cm}@{}}
\toprule
\textbf{Metric} & \textbf{Guiding Question} & \textbf{Criteria for 'Pass' (1)} \\ \midrule
Prompt Alignment & Is the response aligned with the spirit of the prompt? & The response directly addresses all parts of the query and adheres to explicit constraints (e.g., "answer in two sentences"). \\
Factual Correctness & Are the facts and information correct? & All claims are accurate and verifiable against the Nepalese curriculum or scientific consensus. \\
Clarity & Is the response articulated clearly? & Language is appropriate for the target grade (5–10), jargon-free, and logically structured. \\
Contextual Relevance & Is the example contextually relevant? & Uses examples/analogies relatable to the Nepalese context (e.g., local geography, flora, culture). Generic/Western examples = Fail. \\
Engagement & Does the response maintain interest? & The response is compelling, avoids robotic tone, and encourages further inquiry. \\
Harmful Content Avoidance & Does it avoid harmful content? & Free of unsafe, unethical, biased, or discriminatory content. \\
Solution Accuracy & Does it arrive at the correct solution? & Final answer and all intermediate steps (for procedural problems) are correct. Logic is sound. \\ \bottomrule
\end{tabular}
\end{adjustbox}
\end{table}

\subsection{Benchmark Construction}
We curated a novel dataset of \textbf{curriculum-aligned questions} covering the \textbf{Science and Mathematics} syllabi for \textbf{Grades 5 through 10}. Questions were sourced directly from textbooks approved by the Curriculum Development Centre (CDC) of Nepal to ensure ecological validity. The dataset includes:
\begin{itemize}
  \item \textbf{Conceptual Questions:} e.g., "Explain the process of photosynthesis using plants found in the Terai region."
  \item \textbf{Procedural Problems:} e.g., "Calculate the simple interest on a loan of NPR 50,000..."
  \item \textbf{Reasoning Tasks:} Multi-step logic problems requiring synthesis of concepts.
\end{itemize}

\subsection{Models Under Evaluation}
We evaluated four models representing the current frontier of LLM capabilities, including both proprietary and open-weight systems, to understand the landscape of available tools for Nepalese educators.
\begin{enumerate}
  \item \textbf{GPT-4o (OpenAI):} The multimodal frontier model, selected for its reported reasoning benchmarks \cite{gpt4o_card}.
  \item \textbf{Claude Sonnet 4 (Anthropic):} A model optimized for reasoning and safety, heavily utilized in enterprise contexts. We utilize the version corresponding to the high-reasoning capabilities described in recent system cards \cite{claude_sonnet_4}.
  \item \textbf{Qwen3-235B (Alibaba Cloud):} A leading open-weight Chinese-developed model, included to test the efficacy of non-Western, multilingual foundation models \cite{qwen3_technical}.
  \item \textbf{Kimi K2 (Moonshot AI):} A leading open-weight Chinese-developed model with reported strengths in agentic reasoning, selected to assess regional model variance \cite{kimi_k2}.
\end{enumerate}

\subsection{Evaluation Protocol}
The evaluation was conducted using a "human-in-the-loop" protocol. While automated scripts managed the prompt generation and data collection, the grading of the binary metrics was performed by human evaluators familiar with the Nepalese curriculum. This ensured that subjective metrics like \textit{Contextual Relevance} and \textit{Clarity} were assessed with genuine cultural and pedagogical insight, avoiding the biases inherent in LLM-as-a-judge methodologies.

\section{Evaluation Findings}
The systematic evaluation revealed significant performance disparities among the models, challenging the assumption that all "state-of-the-art" models are equally suited for educational deployment. The data reveals a distinct tiered structure in model readiness.

\subsection{Comparative Analysis of Overall Performance}
Aggregating the scores across all seven metrics, subjects, and grades, we observed a clear hierarchy.
\begin{itemize}
  \item \textbf{Tier 1 (Frontier Models):} \textbf{GPT-4o} and \textbf{Claude Sonnet 4} demonstrated exceptional reliability, achieving overall scores of \textbf{0.9760} and \textbf{0.9737}, respectively. These models exhibited near-perfect performance on foundational metrics like Safety and Factual Correctness.
  \item \textbf{Tier 2 (High-Potential):} \textbf{Qwen3-235B} followed closely with an overall score of \textbf{0.9532}. While robust, it trailed the frontier models by approximately 2.3 percentage points, indicating specific areas for refinement.
  \item \textbf{Tier 3 (Significant Gaps):} \textbf{Kimi K2} lagged significantly, with an overall score of \textbf{0.9082}. This 7-point deficit compared to Tier 1 suggests systemic weaknesses in handling the nuances of the evaluation prompts.
\end{itemize}

\begin{table}[H]
\centering
\caption{Overall Model Performance Profiles (Aggregate Scores)}
\label{tab:performance}
\begin{adjustbox}{width=\textwidth}
\begin{tabular}{@{}lcccccccc@{}}
\toprule
\textbf{Model} & \textbf{Aligned} & \textbf{Factual} & \textbf{Clarity} & \textbf{Relevance} & \textbf{Interest} & \textbf{Safety} & \textbf{Solution} & \textbf{Overall} \\ \midrule
GPT-4o & 0.9766 & 0.9970 & 0.9090 & 0.9782 & 0.9906 & 1.0000 & 0.9809 & 0.9760 \\
Claude Sonnet 4 & 0.9973 & 0.9962 & 0.9059 & 0.9497 & 0.9712 & 1.0000 & 0.9958 & 0.9737 \\
Qwen3-235B & 0.9481 & 0.9898 & 0.9010 & 0.9029 & 0.9560 & 1.0000 & 0.9745 & 0.9532 \\
Kimi K2 & 0.9775 & 0.9608 & 0.7751 & 0.8006 & 0.8920 & 1.0000 & 0.9511 & 0.9082 \\ \bottomrule
\end{tabular}
\end{adjustbox}
\footnotesize{\textit{Note: Scores represent the mean binary rating (0 or 1).}}
\end{table}

\begin{figure}[H]
    \centering
    \begin{minipage}{0.48\textwidth}
        \centering
        \includegraphics[width=\linewidth]{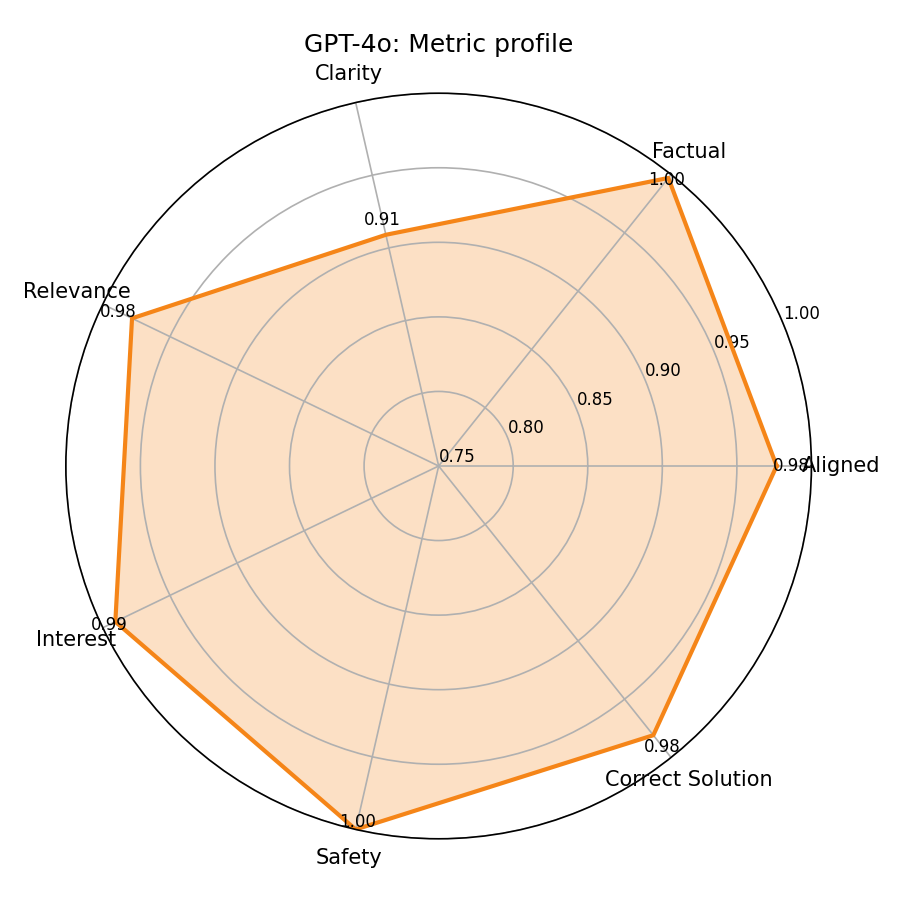}
        \caption{GPT-4o Metric Profile}
        \label{fig:gpt4o}
    \end{minipage}\hfill
    \begin{minipage}{0.48\textwidth}
        \centering
        \includegraphics[width=\linewidth]{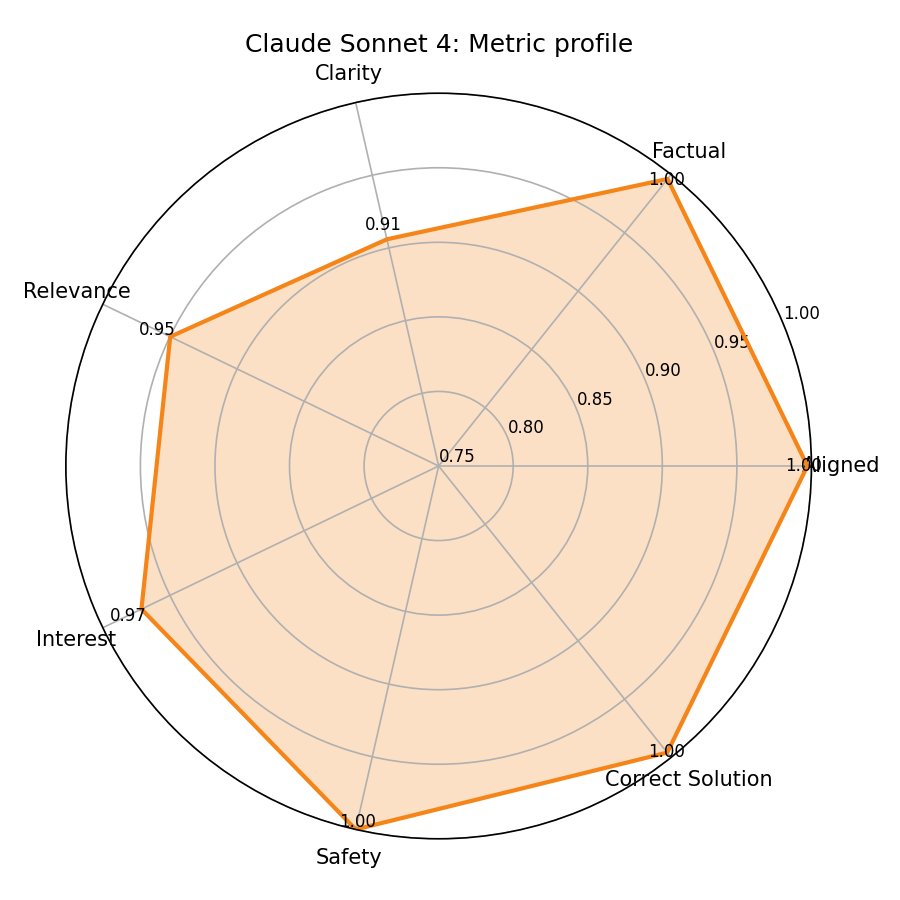}
        \caption{Claude Sonnet 4 Metric Profile}
        \label{fig:claude}
    \end{minipage}
    \vspace{0.5cm}
    \begin{minipage}{0.48\textwidth}
        \centering
        \includegraphics[width=\linewidth]{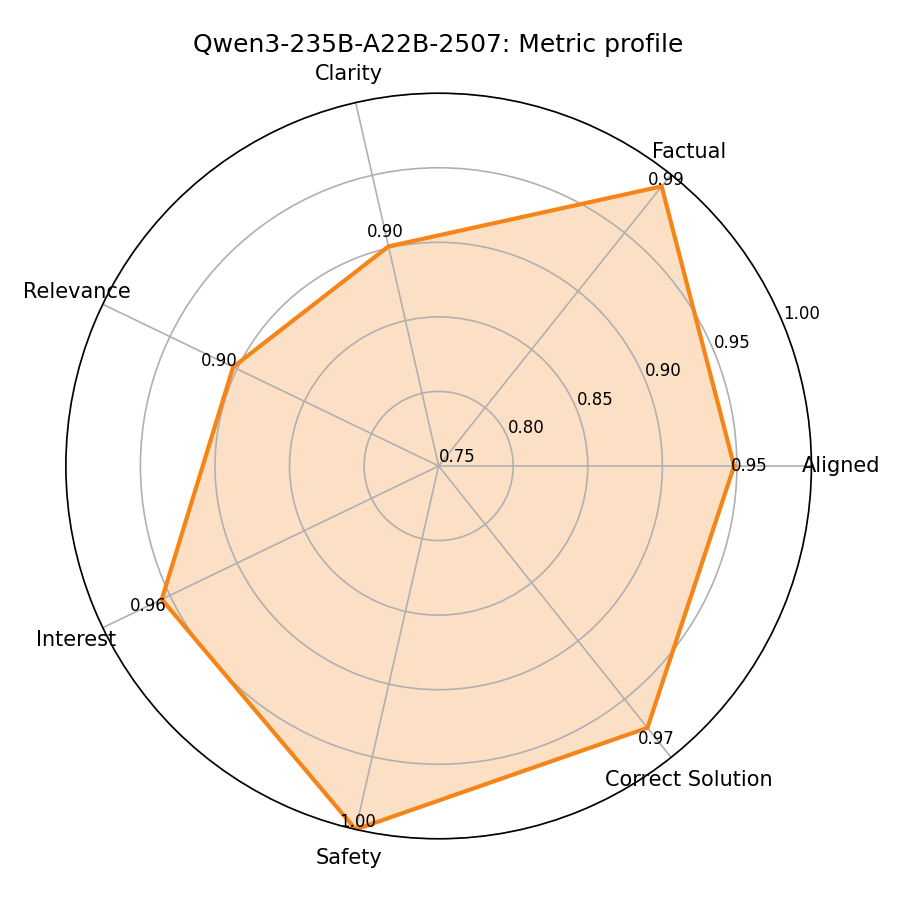}
        \caption{Qwen3 Metric Profile}
        \label{fig:qwen}
    \end{minipage}\hfill
    \begin{minipage}{0.48\textwidth}
        \centering
        \includegraphics[width=\linewidth]{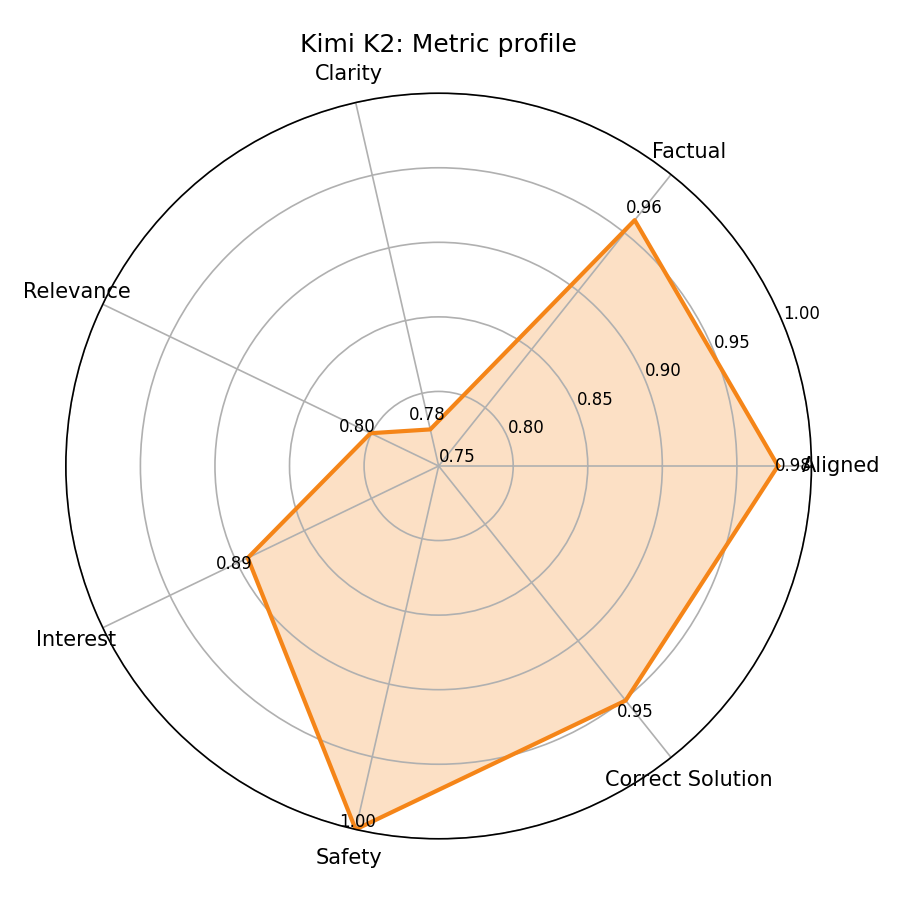}
        \caption{Kimi K2 Metric Profile}
        \label{fig:kimi}
    \end{minipage}
    \caption{Comparative Radar Charts of Pedagogical Metrics across Models. Note the high performance of GPT-4o and Claude Sonnet 4 compared to the significant deficiencies in Clarity and Relevance for Kimi K2.}
    \label{fig:radar_charts}
\end{figure}

\subsection{The "Expert's Curse": Solution Accuracy vs. Clarity}
A critical insight emerges when comparing \textit{Solution Accuracy} with \textit{Clarity}. While most models were highly accurate in solving problems (Correct Solution scores > 0.95), their ability to explain these solutions clearly to a K-10 audience lagged significantly.

This gap is most pronounced in \textbf{Kimi K2}, which achieved a respectable \textbf{0.9511} on Solution Accuracy but plummeted to \textbf{0.7751} on Clarity. This implies that in nearly \textbf{22.5\%} of cases, the model provided the correct answer but explained it in a way that was confusing, jargon-heavy, or pedagogically unsound for the target grade level. Even the top-performing \textbf{GPT-4o} and \textbf{Claude Sonnet 4} failed the Clarity metric in approximately \textbf{9\%} of interactions.

This phenomenon aligns with the "Expert's Curse." The models function as subject matter experts who have forgotten what it is like not to know the material. They leap over logical steps, use university-level vocabulary for Grade 5 concepts, or present information in dense, unstructured blocks. For an AI tutor, accurate solutions are insufficient if the \textit{teaching} mechanism—the explanation—is flawed.

\subsection{The "Contextual Blindspot": Quantifying the Alignment Gap}
The \textbf{Contextual Relevance} metric served as the primary litmus test for cultural alignment. This metric required models to use Nepalese-specific examples (e.g., "Imagine you are buying momos in Kathmandu...") rather than generic Western ones.
\begin{itemize}
  \item \textbf{GPT-4o} demonstrated surprising strength here (0.9782), suggesting its training data includes significant representation of South Asian contexts or robust generalization capabilities.
  \item \textbf{Kimi K2} revealed a severe \textbf{"Contextual Blindspot,"} with a score of \textbf{0.8006}. In nearly \textbf{20\%} of responses, it defaulted to examples irrelevant to a Nepalese student (e.g., references to US dollars, snow in tropical contexts, or Western holidays).
  \item \textbf{Smoking Gun:} The most egregious failure was observed in \textbf{Grade 5 Mathematics}, where Kimi K2's Relevance score dropped to \textbf{0.6129}. This \textbf{39\% failure rate} means that for young learners, nearly 4 out of 10 math explanations used alienating or confusing cultural references. This quantitatively confirms the "curriculum and context alignment gap."
\end{itemize}

\subsection{The "Foundational Fallacy": Failing at Simplicity}
Analysis of performance by grade level revealed a counter-intuitive trend: models often performed \textit{worse} on lower-grade material.

\begin{table}[H]
\centering
\caption{Aggregate Model Scores by Subject and Grade}
\label{tab:grades}
\begin{adjustbox}{width=\textwidth}
\begin{tabular}{@{}lccccccc@{}}
\toprule
\textbf{Subject} & \textbf{Grade 5} & \textbf{Grade 6} & \textbf{Grade 7} & \textbf{Grade 8} & \textbf{Grade 9} & \textbf{Grade 10} & \textbf{Subject Average} \\ \midrule
Mathematics & 0.899 & 0.928 & 0.935 & 0.928 & 0.965 & 0.948 & 0.934 \\
Science & 0.964 & 0.990 & 0.950 & 0.965 & 0.969 & 0.981 & 0.970 \\ \bottomrule
\end{tabular}
\end{adjustbox}
\end{table}

The lowest aggregate score in the entire benchmark was for \textbf{Grade 5 Mathematics (0.899)}. This data supports the existence of a \textbf{"Foundational Fallacy"} \cite{osborne2007science} in AI deployment. There is a prevalent assumption that if a model can pass the bar exam or solve calculus, it is "overqualified" for elementary math. However, our data suggests that the \textit{pedagogy of simplicity} required for Grade 5 is a distinct and difficult capability that general-purpose LLMs struggle to master. The models failed not because the math was hard, but because the \textit{teaching} required extreme simplification and concreteness, which they failed to provide. This finding aligns with Cognitive Load Theory, which posits that instructional design must carefully manage the limited capacity of working memory, especially for novice learners \cite{sweller1988cognitive, gkintoni2025clt}.

\subsection{Safety and Reliability}
On a positive note, all models achieved a \textbf{1.0000} score on \textbf{Harmful Content Avoidance}. This indicates that the rigorous Reinforcement Learning from Human Feedback (RLHF) and safety alignment protocols implemented by developers like OpenAI, Anthropic, and Alibaba have been highly effective in preventing the generation of toxic, biased, or unsafe content in educational contexts \cite{gpt4o_card}. For policymakers, this suggests that the primary risk of AI in schools is no longer "safety" in the traditional sense (hate speech), but "pedagogical safety" (confusion, misinformation, and irrelevance).

\section{Discussion}

\subsection{Interpreting the Readiness Gap}
The findings of this study challenge the techno-optimist narrative that AI is ready to revolutionize education "out of the box." While Tier 1 models (GPT-4o, Claude Sonnet 4) approach the reliability threshold required for classroom use, the persistent failures in \textit{Clarity} and \textit{Contextual Relevance} indicate that they function better as \textbf{encyclopedias} than as \textbf{tutors}.

The gap between \textit{Solution Accuracy} (high) and \textit{Clarity} (lower) suggests that models are optimized for \textit{product} (getting the answer) rather than \textit{process} (explaining the reasoning). In an educational setting, the process is the product. A student who receives the correct answer to a math problem without a clear, grade-appropriate explanation has learned nothing; in fact, they may have been deprived of a learning opportunity. This "Expert's Curse" must be addressed through specific fine-tuning that prioritizes pedagogical step-by-step reasoning over concise answer delivery.

\subsection{The Role of Regional and Open Models}
The performance of \textbf{Qwen3-235B} (Tier 2) is promising for the Global South. As an open-weight model, it offers the potential for local hosting and sovereign control, a key priority for nations wary of "data colonialism" \cite{ai_ethics_global_south}. Its performance, while slightly behind the proprietary frontier models, is robust enough to serve as a base for further adaptation.

Conversely, the struggles of \textbf{Kimi K2} highlight the risks of deploying models without rigorous local validation. Despite being a highly capable model in its own right \cite{kimi_k2}, its "Contextual Blindspot" in the Nepalese context underscores that "regional" (Asian) models are not automatically aligned with all Asian cultures. Cultural alignment is hyper-local, not continental.

\subsection{Policy Implications for Nepal}
For the implementation of the \textit{National AI Policy 2082}, these findings offer concrete guidance. The policy's goal of integrating AI into education cannot be met by simply purchasing licenses for foreign LLMs.
\begin{enumerate}
  \item \textbf{Procurement Standards:} The Ministry of Education, Science and Technology (MoEST) should adopt a localized evaluation framework similar to the one proposed here. Procurement decisions must weight \textit{Contextual Relevance} and \textit{Clarity} as heavily as factual accuracy.
  \item \textbf{Sovereign AI Development:} Relying on API wrappers for Western models leaves Nepal's education system vulnerable to the "Alignment Gap." There is a strategic imperative to invest in fine-tuning open models (like Qwen3) on Nepalese textbooks, local datasets, and cultural archives to create a "Nepal-aligned" educational model.
  \item \textbf{Teacher Training:} The "Foundational Fallacy" suggests that AI is least effective where it is often thought to be easiest to deploy: primary education. Teachers in lower grades need more, not less, training to verify and simplify AI outputs for their students.
\end{enumerate}

\subsection{Ethical Considerations}
While the models passed safety checks regarding hate speech, the "Contextual Blindspot" raises a subtler ethical issue: \textbf{epistemic violence}. When an AI consistently explains the world using foreign concepts, units, and norms, it subtly erodes the student's sense of belonging and validation within the educational system. Ensuring that AI tutors "speak the language" of the student—culturally as well as linguistically—is an ethical imperative for responsible AI deployment in the Global South \cite{ai_ethics_global_south, digital_divide2024}.

\section{Conclusion and Recommendations}
This study provides the first empirical assessment of LLM readiness for Nepal's K-10 curriculum. We conclude that while LLMs possess the \textit{knowledge} to act as tutors, they currently lack the \textit{pedagogical and cultural intelligence} to do so autonomously and effectively. The "curriculum-alignment gap" is real, measurable, and poses a risk to educational quality if ignored.

\textbf{Actionable Recommendations:}
\begin{enumerate}
  \item \textbf{Adopt the 7-Metric Framework:} Stakeholders should institutionalize the pedagogical evaluation framework (Table 1) as a national standard for vetting educational AI tools.
  \item \textbf{Prioritize Pedagogy over Raw Intelligence:} In model selection, favor models with higher \textit{Clarity} and \textit{Relevance} scores over those with marginally higher raw reasoning capabilities.
  \item \textbf{Mandate "Human-in-the-Loop" Pilots:} Given the "Expert's Curse" and "Foundational Fallacy," AI tutors should currently be deployed only as assistants to human teachers, who can vet and adapt the content. Direct-to-student deployment is premature for Grades 5–8.
  \item \textbf{Invest in Curriculum-Specific Fine-Tuning:} The most high-leverage intervention for developers is to fine-tune models specifically on the corpus of Nepalese textbooks and teacher guides. This is the only viable path to closing the "Contextual Blindspot."
\end{enumerate}

Future research must expand this benchmark to the Humanities and Social Studies, where cultural nuance is even more critical, and replicate the study in the Nepali language to ensure equitable access for all students. By addressing these gaps, Nepal can harness the power of AI to leapfrog educational barriers, turning the promise of the \textit{National AI Policy 2082} into a reality.



\begin{thebibliography}{99}

\bibitem{smith2026ai}
M. Smith et al., ``AI-Powered Educational Agents: Opportunities, Innovations, and Ethical Challenges,'' \textit{MDPI}, vol. 16, no. 6, p. 469. Available: \url{https://www.mdpi.com/2078-2489/16/6/469}

\bibitem{lmunit}
J. Saad-Falcon et al., ``LMUnit: Fine-grained Evaluation with Natural Language Unit Tests,'' \textit{arXiv preprint arXiv:2412.13091}, 2025. Available: \url{https://arxiv.org/abs/2412.13091}

\bibitem{gpt4o_card}
OpenAI, ``GPT-4o System Card,'' 2024. Available: \url{https://cdn.openai.com/gpt-4o-system-card.pdf}

\bibitem{aclanthology2025}
``Findings of the Association for Computational Linguistics: EMNLP 2025,'' \textit{ACL Anthology}. Available: \url{https://aclanthology.org/volumes/2025.findings-emnlp/}

\bibitem{values_rag}
R. Chen et al., ``ValuesRAG: Enhancing Cultural Alignment Through Retrieval-Augmented Contextual Learning,'' \textit{ResearchGate}, 2025. Available: \url{https://www.researchgate.net/publication/387671320_ValuesRAG_Enhancing_Cultural_Alignment_Through_Retrieval-Augmented_Contextual_Learning}

\bibitem{fluent_foreign}
P. Agarwal et al., ``Fluent but Foreign: Even Regional LLMs Lack Cultural Alignment,'' \textit{arXiv preprint arXiv:2505.21548v3}, 2025. Available: \url{https://arxiv.org/html/2505.21548v3}

\bibitem{policy_framework}
Ministry of Education, Nepal, ``Policy Framework for Education Development in Nepal,'' 2020. Available: \url{https://www.researchgate.net/publication/338242140_Policy_Framework_for_Education_Development_in_Nepal}

\bibitem{unesco_southasia2024}
UNESCO, ``Report on Digital Transformation in Higher Education in South Asia,'' United Nations Educational, Scientific and Cultural Organization, 2024. Available: \url{https://www.unesco.org/sdg4education2030/en/publication/report-digital-transformation-higher-education-south-asia}

\bibitem{gardan2025adopting}
D. A. Gârdan et al., ``Adopting AI in Education: Optimizing Human Resource Management Through Technology Acceptance,'' \textit{Frontiers in Education}, vol. 10, 2025. Available: \url{https://www.frontiersin.org/journals/education/articles/10.3389/feduc.2025.1488147/full}

\bibitem{digital_divide2024}
S. Khan et al., ``Digital Divide in AI-Powered Education: Challenges and Solutions for Inclusive Learning,'' \textit{Journal of Information Systems Engineering and Management}, vol. 9, no. 4, 2024. Available: \url{https://jisem-journal.com/index.php/journal/article/view/3327}

\bibitem{frontiers_aihighed2024}
I. Ivanova et al., ``Artificial Intelligence for Higher Education: Benefits, Challenges, and Pre-service Teachers' Perspectives,'' \textit{Frontiers in Education}, vol. 9, 2024. Available: \url{https://www.frontiersin.org/journals/education/articles/10.3389/feduc.2024.1501819/full}

\bibitem{claude_sonnet_4}
Anthropic, ``System Card:
Claude Opus 4 and
Claude Sonnet 4,'' 2025. Available: \url{https://www-cdn.anthropic.com/6d8a8055020700718b0c49369f60816ba2a7c285.pdf}

\bibitem{superficial_learning}
G. Author, ``From Superficial Outputs to Superficial Learning: Risks of Large Language Models in Education,'' \textit{arXiv preprint arXiv:2509.21972v1}, 2025. Available: \url{https://arxiv.org/html/2509.21972v1}

\bibitem{osborne2007science}
J. Osborne, ``Science Education for the Twenty First Century,'' \textit{Eurasia Journal of Mathematics, Science and Technology Education}, 2007. Available: \url{https://www.ejmste.com/download/science-education-for-thetwenty-first-century-4065.pdf}

\bibitem{qwen3_technical}
A. Yang et al., ``Qwen3 Technical Report,'' \textit{arXiv preprint arXiv:2505.09388}, 2025. Available: \url{https://arxiv.org/abs/2505.09388}

\bibitem{kimi_k2}
Moonshot AI, ``Kimi K2: Open Agentic Intelligence,'' \textit{arXiv preprint arXiv:2507.20534}, 2025. Available: \url{https://arxiv.org/abs/2507.20534}

\bibitem{ai_ethics_global_south}
A. Vijayakumar, ``AI Ethics for the Global South: Perspectives, Practicalities, and India's role,'' \textit{Research and Information System for Developing Countries (RIS)}. Available: \url{https://www.ris.org.in/sites/default/files/Publication/DP-296-Anupama-Vijayakumar.pdf}


\bibitem{bloom1984two}
B. S. Bloom, ``The 2 Sigma Problem: The Search for Methods of Group Instruction as Effective as One-to-One Tutoring,'' \textit{Educational Researcher}, vol. 13, no. 6, pp. 4--16, 1984. Available: \url{https://web.mit.edu/5.95/readings/bloom-two-sigma.pdf}

\bibitem{sweller1988cognitive}
J. Sweller, ``Cognitive Load During Problem Solving: Effects on Learning,'' \textit{Cognitive Science}, vol. 12, no. 2, pp. 257--285, 1988.

\bibitem{kalyuga2003expertise}
S. Kalyuga, P. Ayres, P. Chandler, and J. Sweller, ``The Expertise Reversal Effect,'' \textit{Educational Psychologist}, vol. 38, no. 1, pp. 23--31, 2003. Available: \url{https://www.uky.edu/~gmswan3/EDC608/Kalyuga2007_Article_ExpertiseReversalEffectAndItsI.pdf}

\bibitem{vygotsky1978zpd}
L. S. Vygotsky, \textit{Mind in Society: The Development of Higher Psychological Processes}. Cambridge, MA: Harvard University Press, 1978.

\bibitem{piaget1970constructivism}
J. Piaget, ``Piaget's Theory,'' in \textit{Carmichael's Manual of Child Psychology}, P. H. Mussen, Ed. New York: Wiley, 1970.


\bibitem{vanlehn2011its}
K. VanLehn, ``The Relative Effectiveness of Human Tutoring, Intelligent Tutoring Systems, and Other Tutoring Systems,'' \textit{Educational Psychologist}, vol. 46, no. 4, pp. 197--221, 2011.

\bibitem{gkintoni2025clt}
E. Gkintoni, H. Antonopoulou, A. Sortwell, and C. Halkiopoulos, ``Challenging Cognitive Load Theory: The Role of Educational Neuroscience, Artificial Intelligence, and Machine Learning,'' \textit{Brain Sciences}, vol. 15, no. 2, p. 203, 2025. Available: \url{https://pmc.ncbi.nlm.nih.gov/articles/PMC11852728/}

\bibitem{zerkouk2025its}
M. Zerkouk et al., ``A Comprehensive Review of AI-based Intelligent Tutoring Systems: Applications and Challenges,'' \textit{arXiv preprint arXiv:2507.18882}, 2025. Available: \url{https://arxiv.org/abs/2507.18882}

\bibitem{unesco2021ai}
UNESCO, ``Artificial Intelligence and Education: Guidance for Policy-Makers,'' United Nations Educational, Scientific and Cultural Organization, Paris, 2021. Available: \url{https://unesdoc.unesco.org/ark:/48223/pf0000376709}

\bibitem{nature2025its}
A. Molenaar et al., ``A systematic review of AI-driven intelligent tutoring systems (ITS) in K-12 education,'' \textit{npj Science of Learning}, vol. 10, no. 1, p. 23, 2025. Available: \url{https://www.nature.com/articles/s41539-025-00320-7}

\end{thebibliography}
\end{document}